\documentclass[10pt,a4paper,pdflatex]{article}
\pdfoutput=1
\usepackage[utf8x]{inputenc}
\usepackage{ucs}
\usepackage[english]{babel}
\usepackage[T1]{fontenc}
\usepackage{amsmath}
\usepackage{amsfonts}
\usepackage{amssymb}
\usepackage{ifpdf}
\usepackage{subfigure}
\ifpdf
\usepackage[pdftex]{graphicx}
\else
\usepackage{graphicx}
\fi
\usepackage{multimedia}
\usepackage{pict2e}
\usepackage{xcolor}
\usepackage{array}

\usepackage{authblk}

\title{Scaling properties of viscous fingering}

\author[1,2]{Bertrand Lagr\'{e}e\thanks{blagree@dalembert.upmc.fr}}
\author[2]{St\'{e}phane Zaleski}
\author[1,3]{Igor Bondino}
\author[2]{Christophe Josserand}
\author[2,4]{St\'{e}phane Popinet}
\affil[1]{TOTAL~SA, Courbevoie, France}
\affil[2]{Sorbonne Universit\'{e}s, UPMC Univ Paris 06, CNRS, UMR 7190, Institut Jean le Rond d'Alembert, F-75005, Paris, France}
\affil[3]{Centre Scientifique et Technique Jean F\'{e}ger, Pau, France}
\affil[4]{National Institute of Water and Atmospheric Research, Wellington, New Zealand}

\begin{document}
\maketitle

\begin{abstract}
  We present a study of viscous fingering using the Volume Of Fluid method and a central injection
  geometry, assuming a Laplacian field and a simple surface tension law. As in experiments we see
  branched structures resulting from the Saffman-Taylor instability. We find that the area $A$ of a
  viscous-fingering cluster varies as a simple power law $A \sim L^\alpha$ of its interface length $L$. 
  Our results are compared to previously published simulations in which 
  the viscosity of the invading fluid is vanishing. We find differences in exponent $\alpha$ and
  in the appearance of detached droplets and bubbles. 
\end{abstract}

Saffman-Taylor's instability \cite{ST1958} is the result of the motion of two viscous fluids in the
narrow space between two parallel plates known as a Hele-Shaw cell \cite{HS1898}, for specific viscosity
ratios. Indeed, when a fluid of lower viscosity displaces a fluid with a higher one, the interface between
them becomes unstable and starts to deform. While the pressure field $p$ satisfies the Laplace equation
\cite{Tang1985}, the interface will move according to Darcy's law 
\cite{AR2013}. 
This problem is formally close both to Diffusion-Limited Aggregation (DLA) \cite{WS1981} and viscous
flows in porous media
\cite{
  Whitaker1986a, Whitaker1986b}.

While Saffman and Taylor used a long and narrow rectangular channel, more recent works have considered
radial \cite{Bataille1968, 
  FS2006, LLFPm2009} and wedge \cite{
  CBa1991, AETT1996} geometries, alongside the historical linear one \cite{
  Bensimon1986, RCG1988, Tanveer2000}, both experimentally and numerically.  Because of the typical fingering
observed in DLA  \cite{WS1983, Sander1986}, porous media experiments
\cite{
  CW1985, MFJ1985
} and Hele-Shaw cells \cite{PH1985}, interest has focused on the fractal aspect \cite{
  Sander2011}.

The Hausdorff dimension $D_F$ 
of a fractal set is often determined by considering the density-density correlation function  \cite{GP1983}. Indeed, this correlation
function varies as $r^{-\delta}$ for small values of $r$ in both viscous fingering and DLA in
circular geometry \cite{CMS1998}. It is easy to demonstrate the exponent is closely related to the
dimension of the considered fractal cluster. Sharon \textit{et al.} \cite{SMmCS2003}
noticed 
that 
the fractal dimension of 
Saffman-Taylor fingering obtained with central injection was close to the one of circular DLA (both
measured \cite{CMS1998} and theoretical \cite{DLP2000}).

As in references \cite{FS2006, LLL2007}, we study the development of Saffman-Taylor fingering in a
Hele-Shaw cell with central injection in the presence of surface tension. For this purpose, we focus on
two-dimensional (2D) square domains $\Omega$ of size $L_d$ centered on
$(0,0)$
. The two components of the velocity are $(u,v)$, \textit{i.e.}, $\textbf{u} = u(x,y)
\textbf{e}_\textbf{x} + v(x,y) \textbf{e}_\textbf{y}$. This domain is initially filled with a
high-viscosity fluid ($\mu = 1$). A less viscous fluid ($\mu = M \ll 1$) is injected from the center of
the domain with a constant massflux 
inside a circle of radius $0.8 \ll L_d$. The side boundaries ($x = \pm L_d/2$ and $y = \pm L_d/2$) impose a
quasi-circular "free" outflow condition for the fluids with Neumann boundary conditions for the
velocity: $\partial_x u (\pm L_d/2, y) = \partial_y v (x, \pm L_d/2) = 0$.

The fluids are considered incompressible ($\nabla \cdot \textbf{u} = 0$) and move inside the domain 
according to Darcy's law
:
\begin{equation} \label{eq:2}
\textbf{u} = - (1/\mu) \nabla p.
\end{equation}
Due to fluid incompressibility and equation (\ref{eq:2}), the pressure field $p$ obeys the Laplace equation $\nabla^2 p = 0$ \cite{Tang1985}.

Boundary conditions are given at the interface which hold for the depth-averaged fields\begin{subequations}:
\begin{align}
[\![ \textbf{n} \cdot \textbf{u} ]\!] & = 0,\\
[\![ p ]\!] & = A + \lambda_{\sigma}/R. \label{eq:3b}
\end{align}
Here $[\![ \text{ } ]\!]$ denotes a jump, $R$ is the principal radius of curvature of the projection, onto the plane, of the tip of the meniscus separating the two phases and $\lambda_{\sigma}$ a characteristic lengthscale of the influence of surface tension, multiplied by
some coefficient of order 1 \cite{PH1984}\end{subequations}.

Let $c(x,y,t)$ be a marker function such that $c(x,y,t) = 1$ in fluid 1 (\textit{i.e.} the invading fluid), and $c(x,y,t) = 0$ in fluid 2 (the receding one).
The surface tension is given by $\textbf{f}_\sigma = \kappa \delta_S \textbf{n}$ where $\kappa$ is the
curvature and $\delta_S(x,y) = |\nabla c(x,y,t)|$ at the interface $S$ in the distribution sense. By
considering that \textbf{n} points towards the invading fluid 1, one can write $\textbf{n} = \nabla c /
|\nabla c|$ and $\kappa = - \nabla \cdot \textbf{n} = A + \lambda_{\sigma}/R$ according to equation
(\ref{eq:3b}).

\begin{subequations}To consider surface tension, equation (\ref{eq:2}) should be written as:
\begin{align}
\textbf{u} & = - (1/\mu) \left[ \nabla p + \kappa(x,y,t) \delta_S \textbf{n} \right] \\
& = - (1/\mu) \left[ \nabla p + \left( A + \lambda_{\sigma}/R \right) \nabla c(x,y,t) \right]. \label{eq:5b}
\end{align}
\end{subequations}
Defining $p^{\star} = p + A c$ in equation (\ref{eq:5b}) allows to use only $\lambda_{\sigma}/R$ in
equation (\ref{eq:3b}) as the pressure jump due to surface tension \cite{AR2013}.

\begin{figure*}
\centering
\subfigure[ ]{
\includegraphics[width=0.18\textwidth]{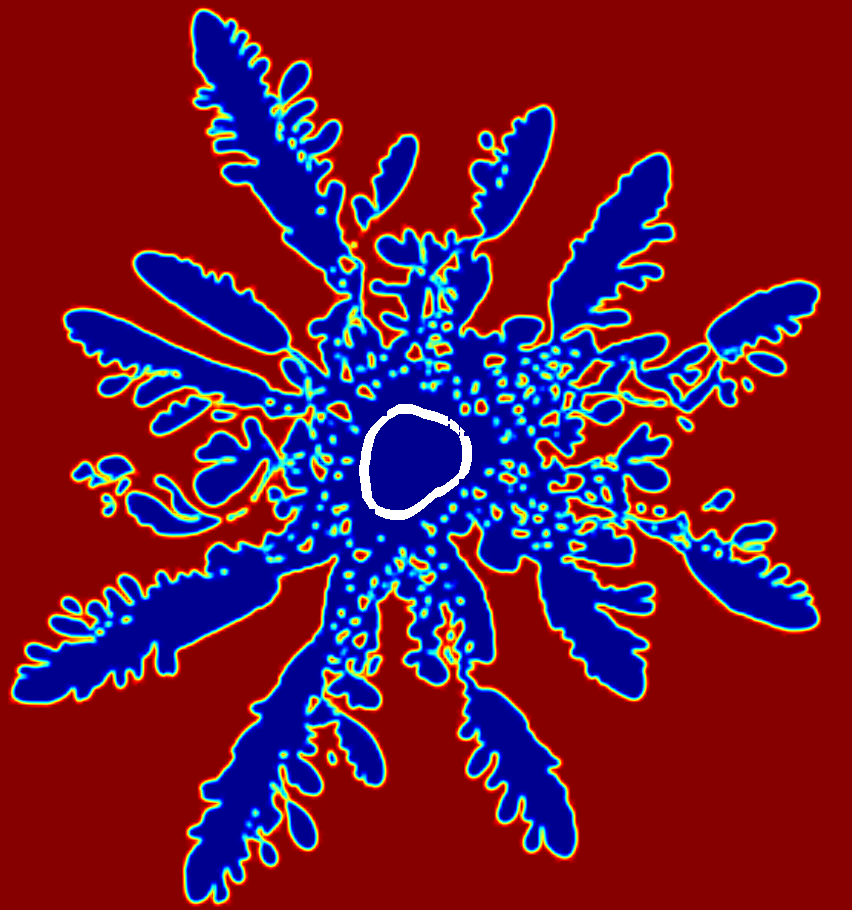}
\label{fig:No2}}
\qquad
\subfigure[ ]{
\includegraphics[width=0.18\textwidth]{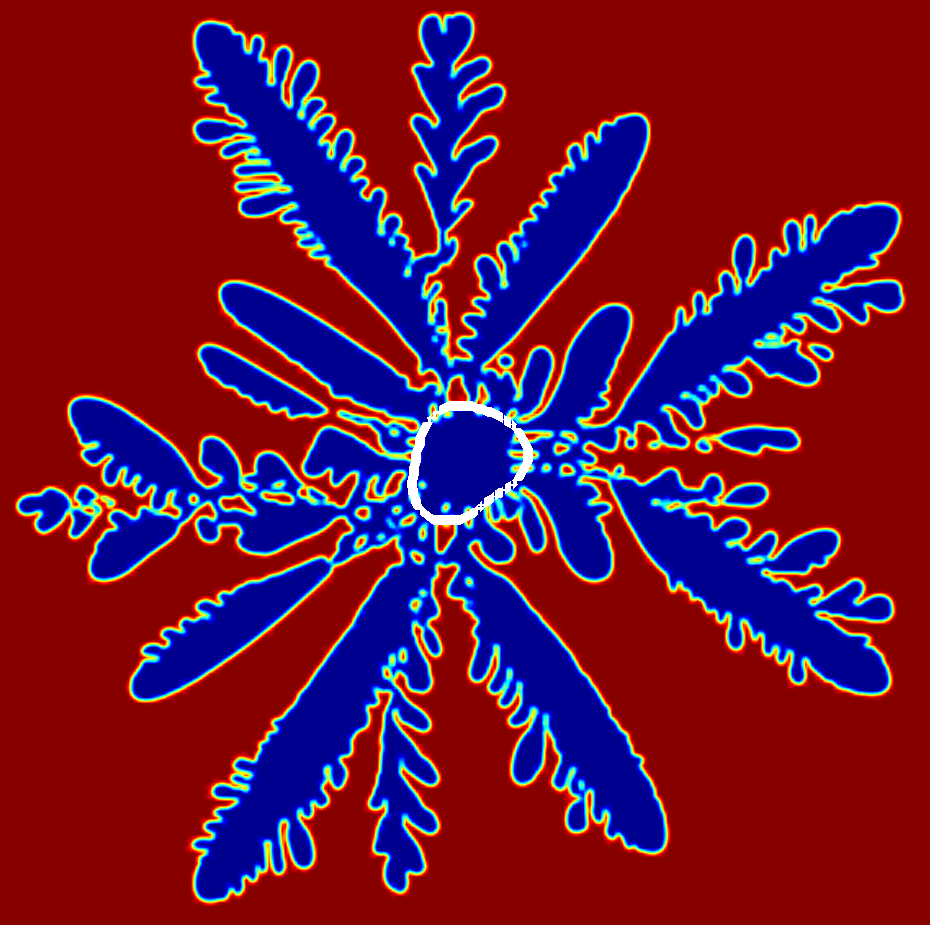}
\label{fig:No3}}
\qquad
\subfigure[ ]{
\includegraphics[width=0.18\textwidth]{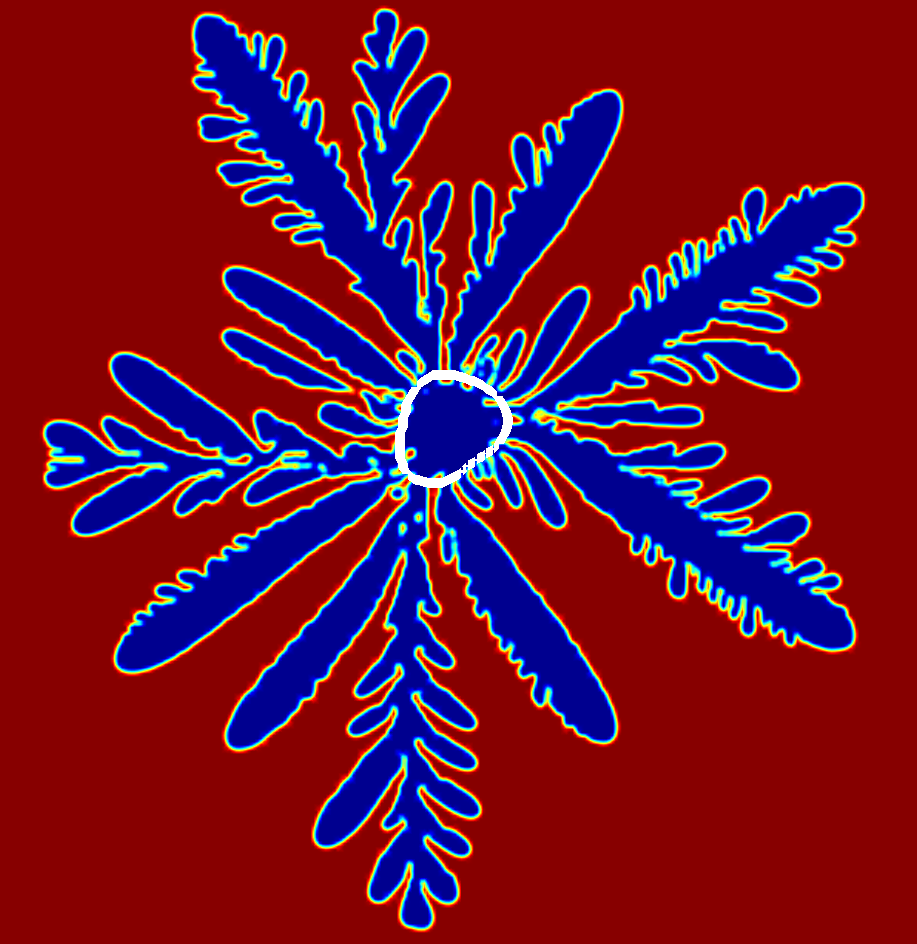}
\label{fig:No4}}

\subfigure[ ]{
\includegraphics[width=0.18\textwidth]{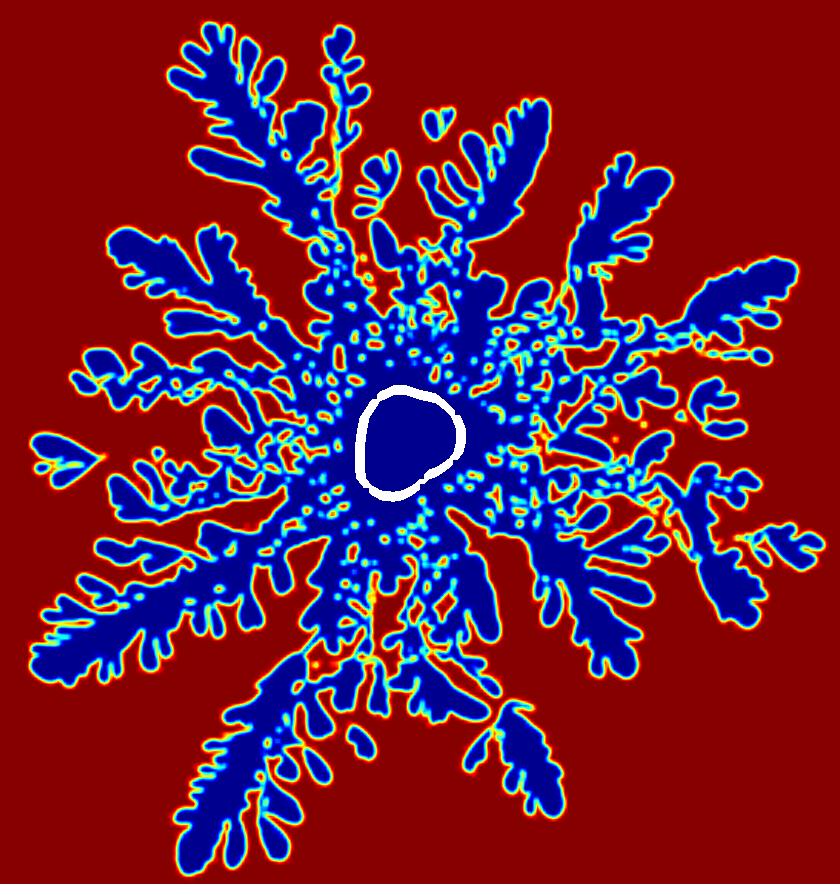}
\label{fig:Add2}}
\qquad
\subfigure[ ]{
\includegraphics[width=0.18\textwidth]{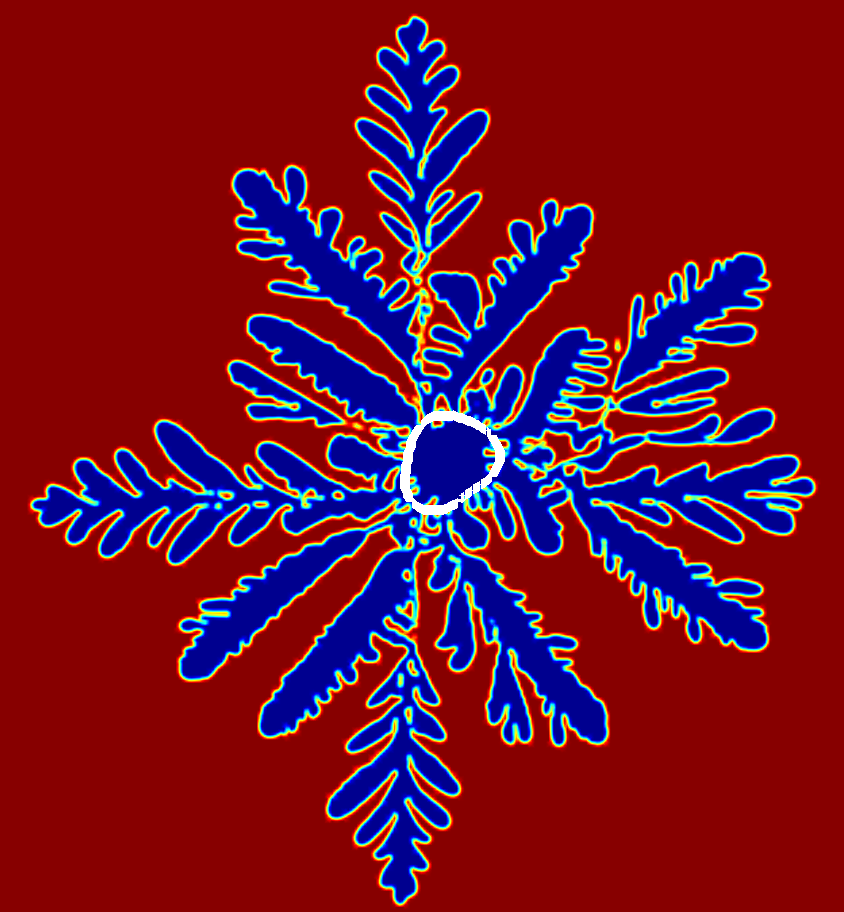}
\label{fig:Add3}}
\qquad
\subfigure[ ]{
\includegraphics[width=0.18\textwidth]{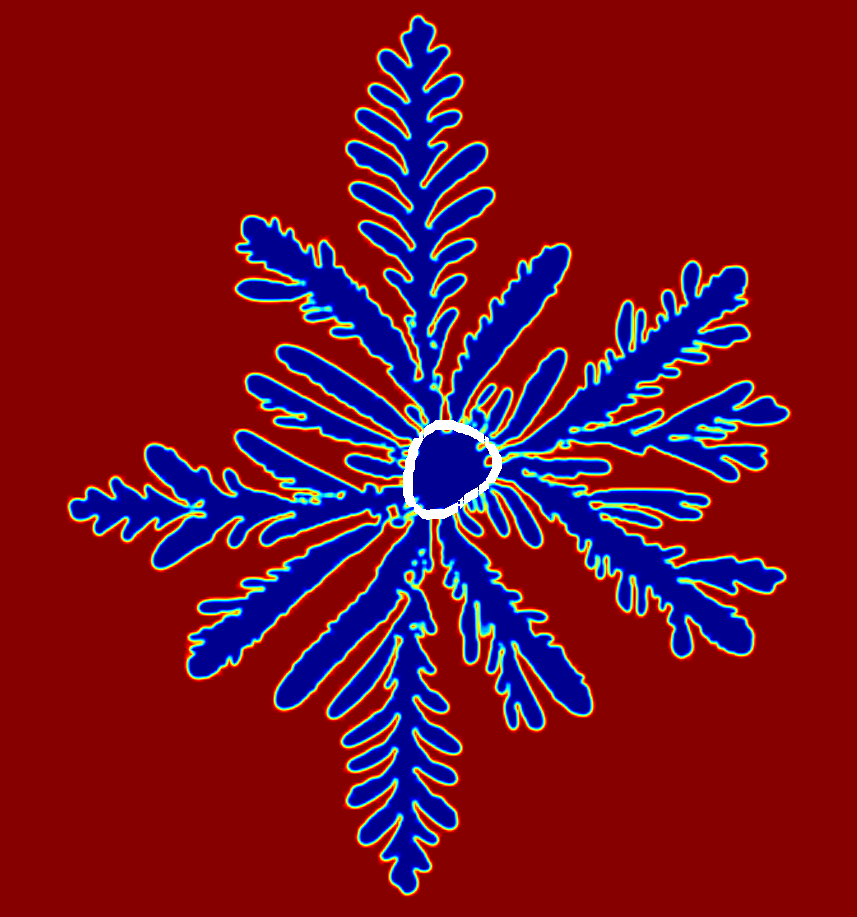}
\label{fig:Add4}}
\caption{
Development of Saffman-Taylor fingering for three different viscosity ratios: $10^{-2}$
  (pictures \subref{fig:No2} and \subref{fig:Add2}), $10^{-3}$ (pictures \subref{fig:No3} and
  \subref{fig:Add3}), $10^{-4}$ (pictures \subref{fig:No4} and \subref{fig:Add4}). The instability is
  generated by mesh-induced noise in the first row, while an isotropic noise is added to the simulations
  presented in the second row. In white, the initial position of the interface.
}
  \label{fig:ST6}
\end{figure*}

Our model is implemented in \textsc{Gerris}, a free-software solver for the solution of incompressible fluid
motion using the finite-volume approach \cite{Popinet2003, Popinet2009}. Gerris uses the Volume-of-Fluid
(VOF) method \cite{TSZ2011} to describe variable-density and/or variable-viscosity two-phase
flows. \begin{subequations}In this method, the Euler equations are written as
  :
\begin{align}
\rho \left( \partial_t \textbf{u} + \textbf{u} \cdot \nabla \textbf{u} \right) & = - \nabla p + \rho \textbf{f}, \label{eq:6a}\\
\partial_t c + \nabla \cdot (c \textbf{u}) & = 0, \\
\rho = \rho(c) & = c \rho_1 + (1 - c) \rho_2.
\end{align}
\end{subequations}

The \textsc{Gerris} flow solver can simulate Darcy flow \cite{AR2013,McBain}. Here we also describe how
to obtain the time dependent problem which was not considered in \cite{AR2013}.
We cancel the advection term of (\ref{eq:6a}) and add a
drag force $\textbf{f} = - \textbf{u}$ : $\partial_t \textbf{u} = - (1/\rho) \nabla p -
\textbf{u}$.  We decompose the numerical solution into an exact and an error term:
$\textbf{u}(t) = \textbf{u}_0 + \textbf{e} (t)$ with $0 = - \nabla p - \rho \textbf{u}_0$ for the exact term.  
The error  \textbf{e} obeys $\partial_t \textbf{e} = - \textbf{e}$ for a steady exact solution
and is $\textbf{e} = - \textbf{C} \exp ( - t )$. The error
will be small after one timestep $\tau$ if $\tau \gg 1$. 
For a time-dependent exact solution $\textbf{u}_0$, the same decomposition shows that the error scales as
$\textbf{u}_0 /T_c$ where $T_c$ is the (dimensionless) characteristic time of the exact solution
and it thus requires that $T_c \gg 1$. The latter condition is achieved by the choice of length and time units. 
Thus the solver produces either a
false-transient iteration towards the steady state or adiabatically follows the time-dependent solution of $\textbf{u} = - (1/\rho) \nabla p$,
which is equivalent to Darcy's law if $\rho = \mu$.

We simulate viscosity ratios $M = \mu_1/\mu_2$ from  $10^{-4}$ to $10^{-2}$. 
Using the pressure field $p^{\star}$ presented alongside equation (\ref{eq:5b}), 
we only consider the planar contribution to the radius of curvature.

The initial data is a slightly perturbed circular interface $r(\theta) = 1 + 1/10 ( \cos (3 \theta) +
\sin (2 \theta))$ \cite{FS2006}. We only report on the early times $r_{\text{max}}/L_d \ll 1$ 
to avoid finite-size effects and mesh-induced anisotropy.  We fix $\lambda_{\sigma} = 1/3$ so that the capillary length scale is somewhat smaller
than the initial radius $r$ and the grid size is $h = 1/60$.  
\begin{table}
\centering
\begin{tabular}{l c c c}
\hline Viscosity ratio & $10^{-2}$ & $10^{-3}$ & $10^{-4}$ \\
\hline
\hline Mesh-induced noise & 1.79 $\rightarrow$ 1.58 & 1.81 $\rightarrow$ 1.60 & 1.69 \\
\hline Added noise & 1.85 $\rightarrow$ 1.56 & 1.68 & 1.70 \\
\hline
\end{tabular}
\caption{
Fractal dimensions of the clusters presented in Figure \ref{fig:ST6}. When two different values are presented, the first one corresponds to the lower radii and the second one to the higher ones.\label{tab:DF}}
\end{table}

\begin{figure*}
\centering
\subfigure[ ]{
\includegraphics[width=0.25\textwidth]{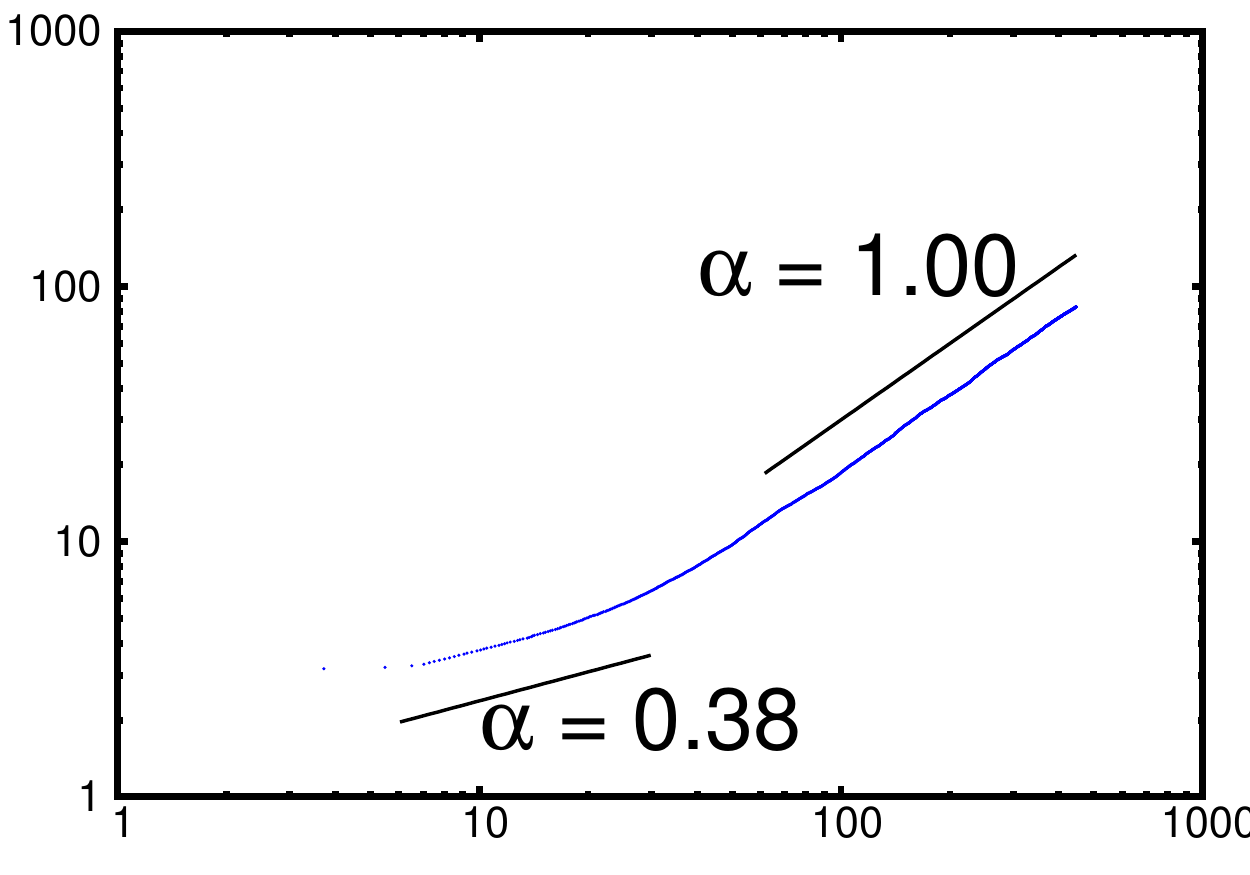}
\label{fig:LA-No2}}
\qquad
\subfigure[ ]{
\includegraphics[width=0.25\textwidth]{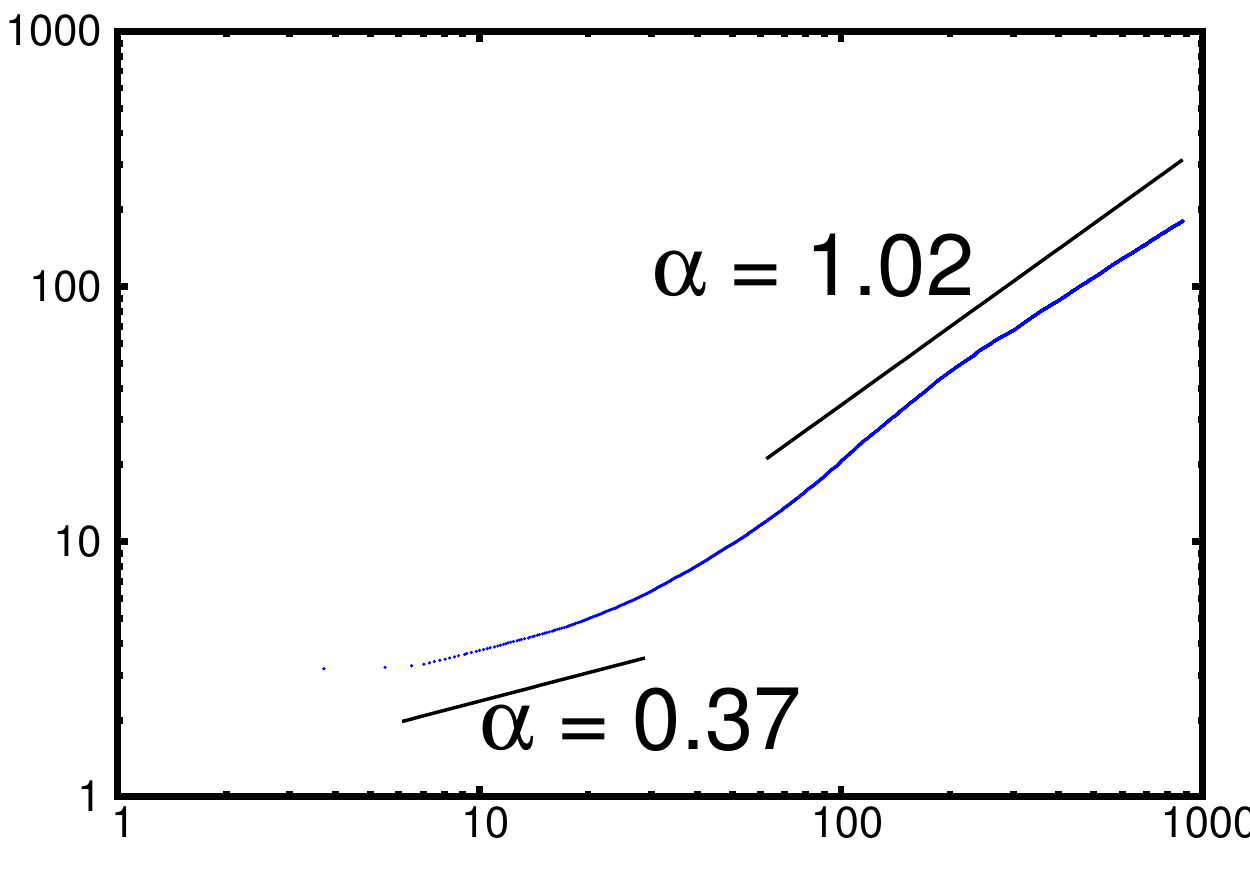}
\label{fig:LA-No3}}
\qquad
\subfigure[ ]{
\includegraphics[width=0.25\textwidth]{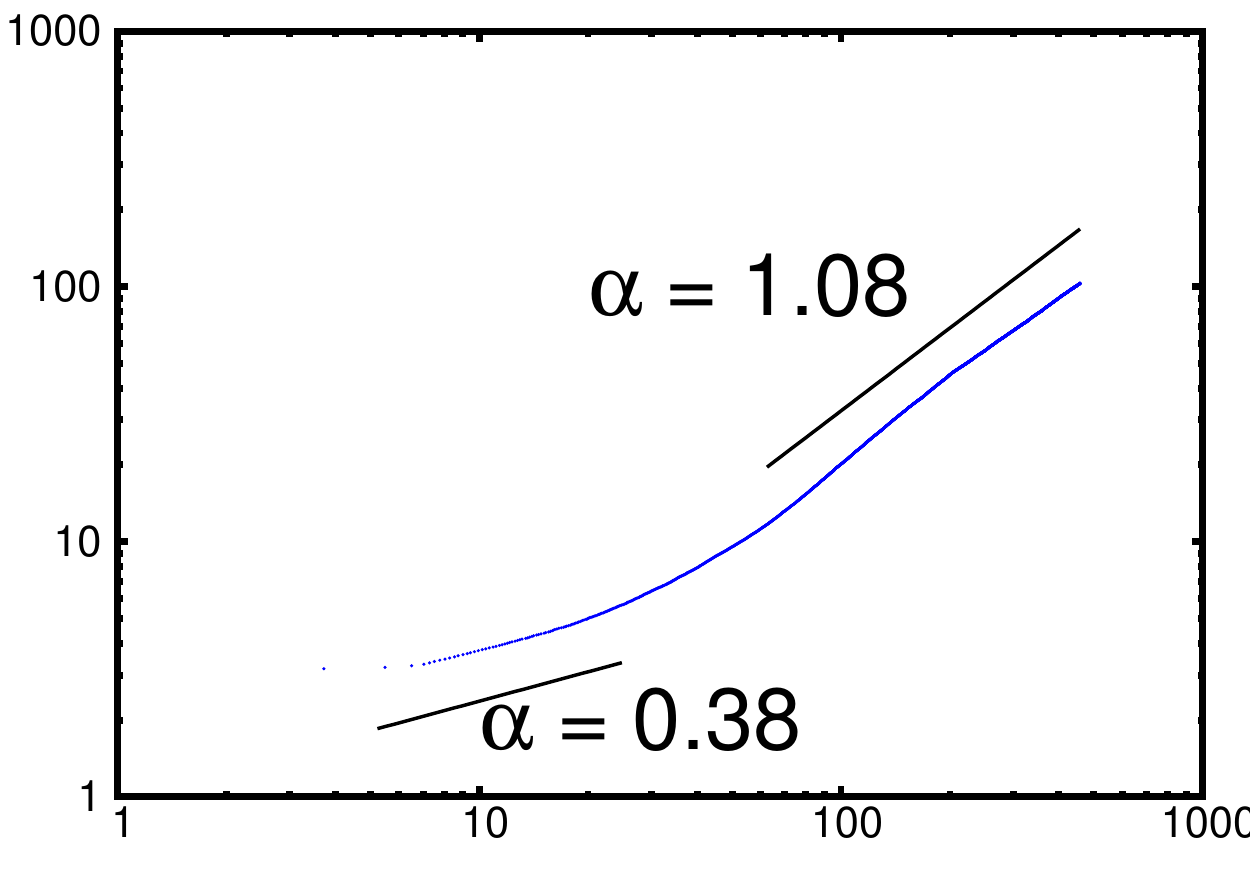}
\label{fig:LA-No4}}

\subfigure[ ]{
\includegraphics[width=0.25\textwidth]{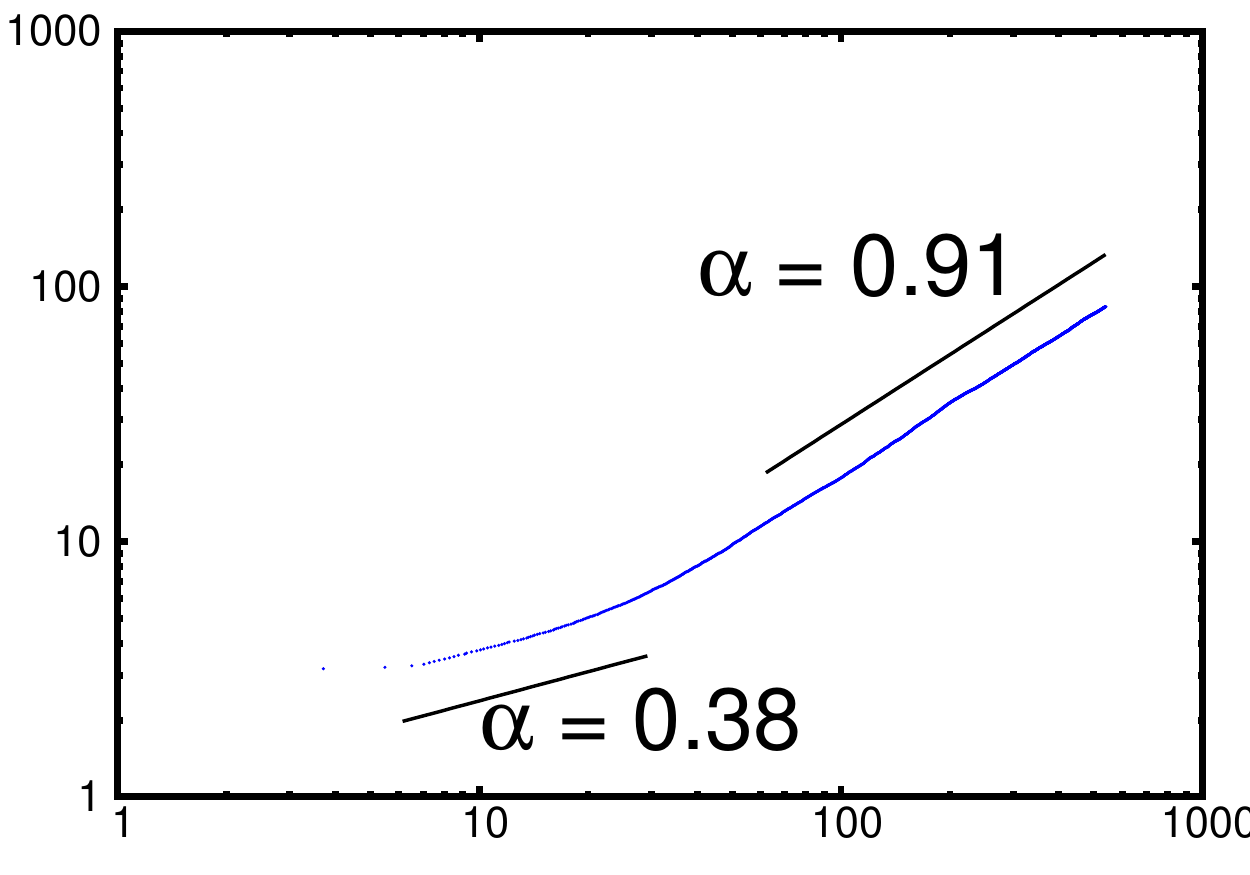}
\label{fig:LA-Add2}}
\qquad
\subfigure[ ]{
\includegraphics[width=0.25\textwidth]{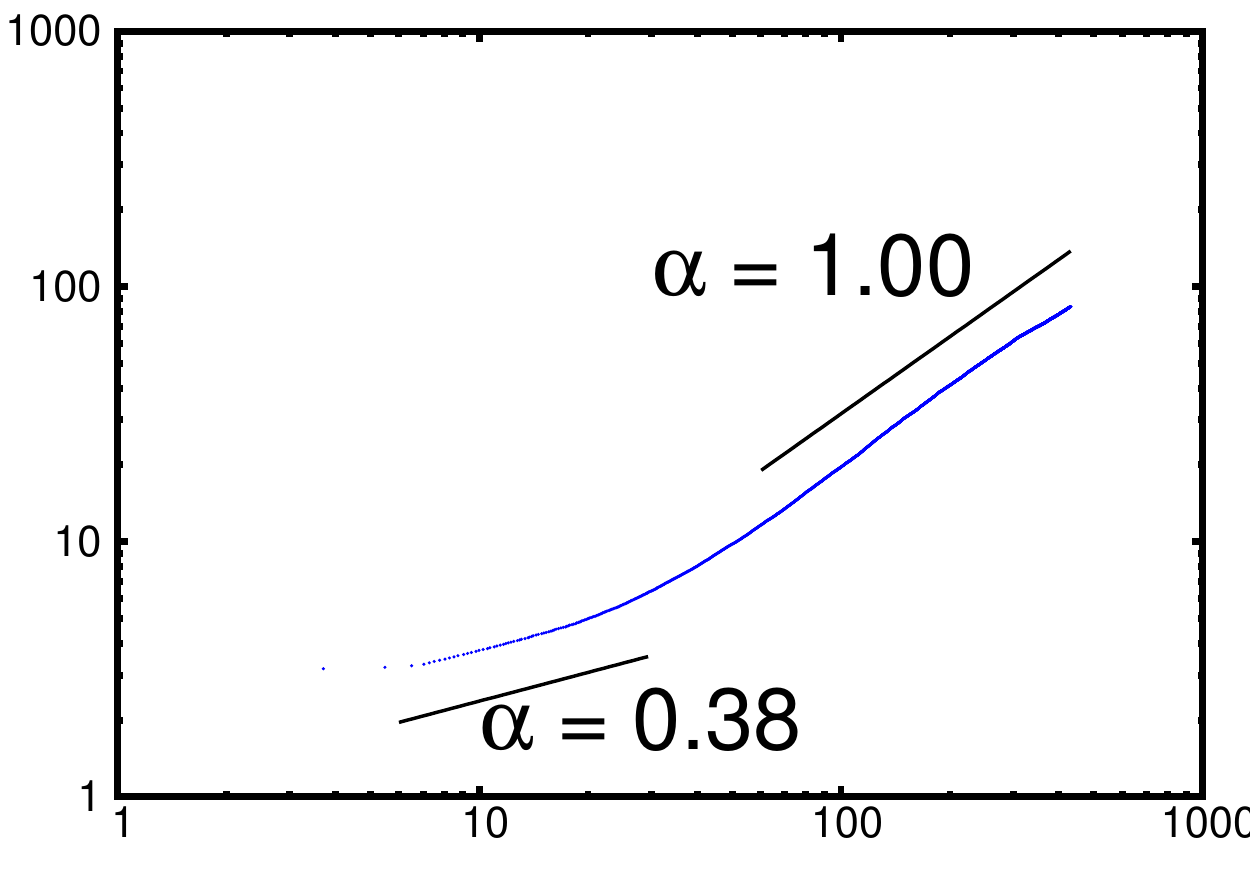}
\label{fig:LA-Add3}}
\qquad
\subfigure[ ]{
\includegraphics[width=0.25\textwidth]{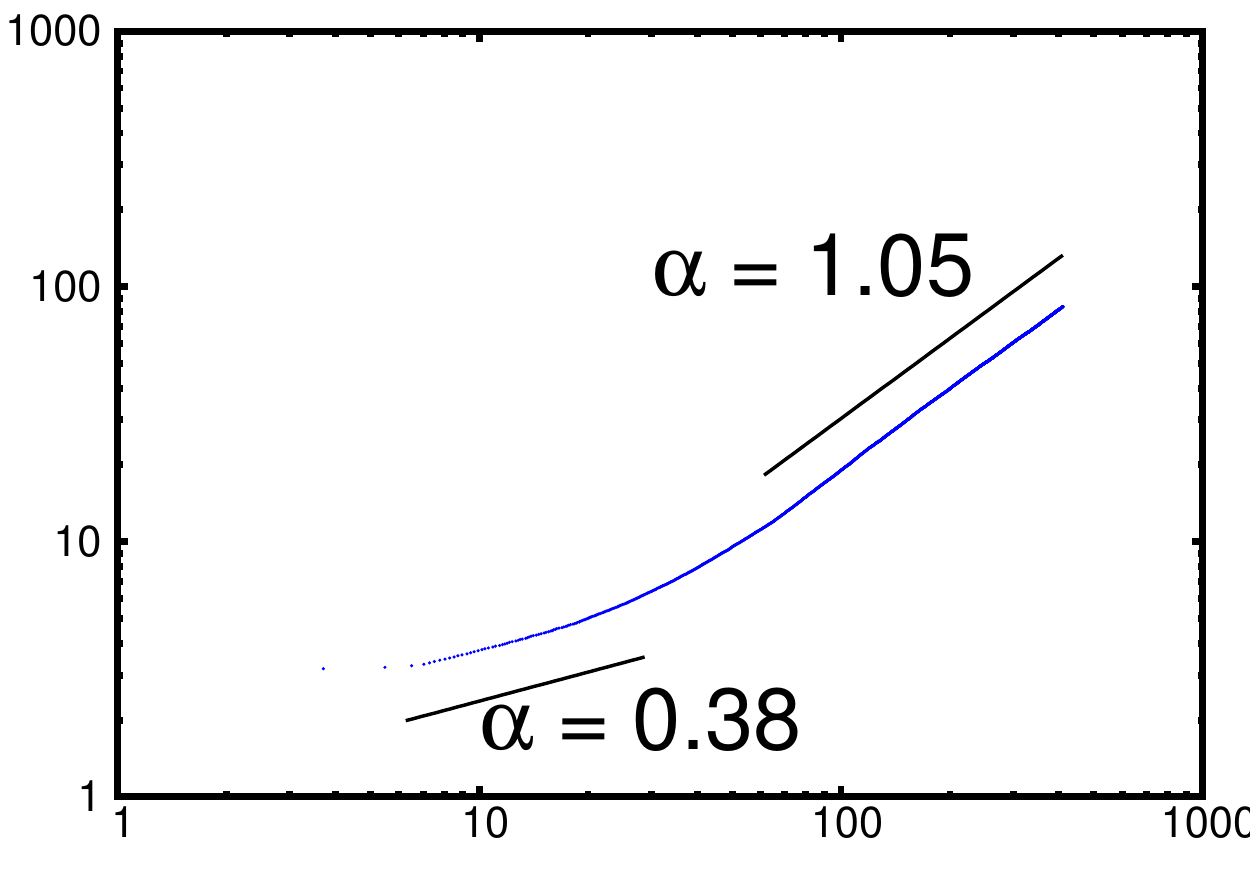}
\label{fig:LA-Add4}}
\caption{
The dimensionless area $A$ of the growing bubble \textit{vs.} the dimensionless length $L$ of
  the interface. The different viscosity ratios and noise origins are presented similarly to Figure
  \ref{fig:ST6}.\label{fig:LA} }
\end{figure*}

In related simulations (invasion from the side of a rectangular domain, to be reported
in a following publication) we found that if care is taken to ensure very low noise levels
fingering disappears and plain, non-branching fingers advance linearly.
 Thus our simulations incorporate two different kind of noise, 
numerical or ``mesh-induced'' noise (due mostly to the finite tolerance on the convergence of the Poisson solver)
or random perturbations of the local viscosity (viscosity added noise). 
The early-time instability amplifies the initial perturbation, then later-time fingerings develop due to
either mesh-induced noise [pictures \ref{fig:No2} to \ref{fig:No4}],
or added noise [pictures \ref{fig:Add2} to \ref{fig:Add4}, strong
enough to cover the still-present mesh-induced noise]. In this last case, the noise is also
axis-dependent, due to the non-isotropic distribution of computational grid points. The viscosity added
noise has standard deviation $[\Delta (1/\mu)]/(<1/\mu>) = 5\%$.
It was added to try to account for the sensitivity of the fingering process to noise \cite{TCC1990}, for instance in porous media.

The error is due to several contributions: numerical uncertainty as stated above, mesh-size effects, finite-size effects (special care was taken to verify they are negligible:
 a specific flow was considered for several domain sizes $L_d$, resulting in convergence with an error lower than 2\%),
  data range used to compute power-law fits, and quality of the fitting process for a specific data range.

At higher viscosity ratios patches are seen to detach from the main
bubble. Moreover, some fingers tend to reconnect at later times, trapping some
high viscosity droplets inside the less viscous main bubble [see pictures \ref{fig:No2} and
\ref{fig:Add2}]. On the contrary, for lower viscosity ratio, the different early-time fingers develop
separately without uniting one with the other [no detaching droplets; pictures \ref{fig:No4} and
\ref{fig:Add4}]. These two different cases result in different fractal-dimension regimes, presented in
Table \ref{tab:DF} (fitting uncertainty always lower than 2\%): at higher viscosity ratio, the inner (lower radii) fractal dimension is higher than
the outer (higher radii) one (around 1.8 \textit{vs.} around 1.6); on the contrary at lower viscosity
ratio, there is only one fractal dimension for the whole bubble (around 1.7).

It can also be noted that the transition from the two-fractal-dimension regime (high viscosity ratio,
see above and Table \ref{tab:DF}) to the one-fractal-dimension one (low viscosity ratio) occurs at higher
viscosity ratio when the noise amplitude is higher. Logically, the higher the noise amplitude, the more
pronounced the resulting fingering (see Figure \ref{fig:ST6}).

When numerically simulating Saffman-Taylor fingering \cite{FS2004, FS2006}, it is common to consider a
constant pressure field $p_0$ inside the less viscous fluid (usually considered equal to zero). This
is equivalent to the assumption of a zero-viscosity invading fluid. In this approximation
and without surface tension, the pressure  is $p_0$ everywhere on the interface. Due to the maximum
principle (a consequence of the open-mapping theorem), the maximum pressure of a Laplacian field
such as $p$ in the domain occupied by the invaded, more viscous fluid 
is only reached at the domain boundary, here the interface. The maximum is thus $p_0$ and
is reached at the interface, with $p<p_0$ everywhere else. Thus the interface can only advance 
towards the more viscous fluid and never recede. This makes the pinch-off of invading fluid bubbles impossible.
The presence of detaching droplets 
is thus a qualitative difference with infinite viscosity contrast models.

In reference \cite{FS2006}, Fast $\&$ Shelley explain that long-time simulations of Saffman-Taylor
fingering reveal an asymptotic scaling regime, where the interface length of the resulting bubble is
related to the bubble area by a power-law relation
: $\text{Area} \sim ( \text{Length} )^{\alpha_{\infty}}$ (in their case, $\alpha_{\infty} = 1.45$). The variation of this
coefficient $\alpha$ is obviously disconnected from that of the fractal dimension $D_F$; indeed, as an
example, one can consider a lone planar dendrite whose length and width grow as $t$ and $t^{\gamma}$
respectively (with $0 < \gamma < 1$), then $D_F = 2$ whatever the value of $\gamma$ whereas $\alpha_{\infty} = 1
+ \gamma$. Approximating fingers by their osculating parabolas one gets $\gamma = 1/2$ and $\alpha_{\infty} = 3/2$. 
In the case of DLA clusters, $\alpha_{\infty} = 1$ whatever the fractal dimension.

Fast $\&$ Shelley \cite{FS2006} also found a short-time scaling with another
coefficient $\alpha_1 = 0.61$. Li \textit{et~al.} \cite{LLL2007} found similar results.

In Figure \ref{fig:LA}, we also find that the interface length is related to the
bubble area by a power-law scaling with two different regimes. However, the coefficients that we obtain
are different: $\alpha_1 \approx 0.38$, and $\alpha_{\infty} \approx 1$,
depending on the viscosity ratio and noise origin. The error, mostly due to mesh-size effects, on all values of both $\alpha_1$ and $\alpha_{\infty}$ is of order 10\%. However 
oscillations in the scaling exponent are inherent to the self-similar character of the fingering 
\cite{SFS1986} and could also account for part of the error.

\begin{figure}[!b]
\centering
\includegraphics[width=0.4\textwidth]{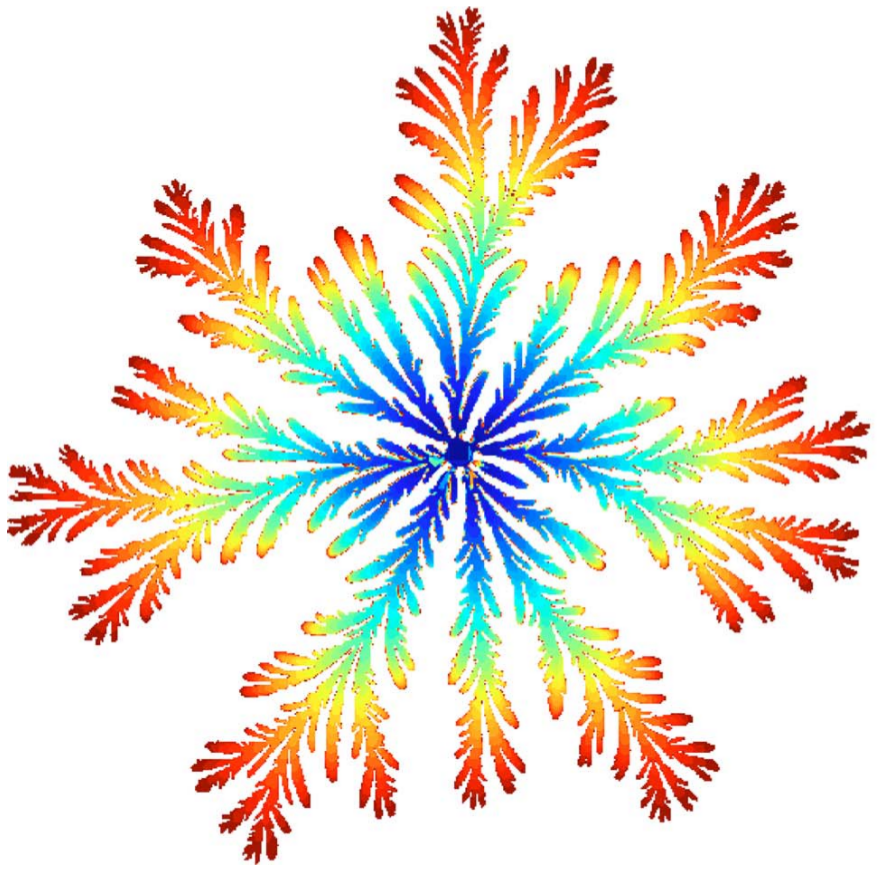}
\caption{
Time evolution of the fractal growth patterns for viscous fingering ($\Delta p = 1.25$
  atm). The colors indicate the ages of the patterns; the oldest (first created) region is blue and the
  youngest is red. From \cite{PS2005}.\label{fig:PS5}}
\end{figure}

It should also be noted that the coefficient $\alpha_{\infty}$ increases slightly when the viscosity ratio decreases,
although we have not enough numerical data to further comment on the significance of this increase. 

Due to the high discrepancy of coefficient values between our results and those of reference
\cite{FS2006}, both Fast $\&$ Shelley's results and ours were compared to the experimental results
presented by Praud $\&$ Swinney in reference \cite{PS2005}. This comparison was realized cautiously,
for, while our results and Fast $\&$ Shelley's were realized at constant massflux, Praud $\&$ Swinney's
injection was realized at constant pressure. The experiments were also realized with the viscosity ratio
$M \sim 5 \cdot 10^{-5}$, \textit{i.e.}, close to what we used for pictures \ref{fig:LA-No4} and
\ref{fig:LA-Add4}.
At $M \sim 5 \cdot 10^{-5}$ in the experiments or $M \sim 10^{-4}$ in the simulation very few 
or even no  detached bubbles or trapped droplets are seen, making this case similar to the $M=0$ one. 

From Figure \ref{fig:PS5}, it was possible to determine $\alpha_{\infty} = 1.14$, with the following fitting error: $\alpha_{\infty} \in [1.11,1.23]$. Our results ($\alpha_{\infty} \in [1.02, 1.08]$ with the fitting uncertainty resulting from data range),
are much closer than those of Fast $\&$ Shelley.

\begin{figure}[!b]
\centering
\includegraphics[width=0.4\textwidth]{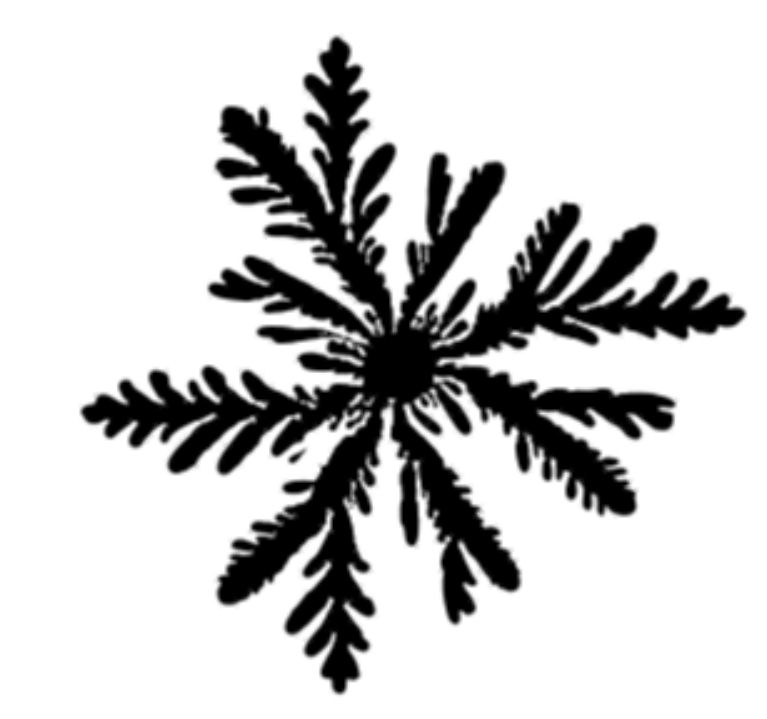}
\caption{
Simulation of the experiment presented in \cite{PS2005} ($r_{\text{max}}/\lambda_{\sigma} = 25$).\label{fig:PS150}}
\end{figure}

In order to completely validate our model, we simulated the experiment presented in Figure
\ref{fig:PS5}. The simulation was realized with a maximum number of authorized computational 
cells of $2^{13}$ in each direction
with the same viscosity ratio as in Figure \ref{fig:PS5}. As there was no blatant difference between the
results of simulations obtained by applying a constant pressure (as in the experiment of Praud $\&$
Swinney) and those with a constant massflux, we present in Figure \ref{fig:PS150} the results realized
with a constant massflux, for which a length $\lambda_{\sigma}$ was defined above.

Though Figures \ref{fig:PS5} and \ref{fig:PS150} look very much alike, it is obvious that our 
fingers begin to align themselves with the principal directions of the mesh: the X-axis, the
Y-axis and both bissectors of the axes.

We nonetheless determined the fractal dimension $D_F = 1.67$ of the resulting cluster, to be compared
with $D_F = 1.69$ in Figure \ref{fig:PS5}. For
all we know, the fractal dimension of a Direct Numerical Simulation of Saffman-Taylor fingering
was never measured beforehand.

As a conclusion, we observe the existence of two different regimes in Saffman-Taylor fingering due to
central injection: one at higher viscosity ratio with the coexistence of two different fractal
dimensions in the resulting cluster, and one at lower viscosity ratio with only one dimension. What is
more, at late times, the area of the resulting bubble varies as the length of the interface to some
power $\alpha_{\infty}$, with $\alpha_{\infty}$ increasing with decreasing viscosity ratio.

\section*{Acknowledgments}

We thank TOTAL for the financial support and permission to publish this study.

\bibliographystyle{unsrt}
\bibliography{arxiv}
\end{document}